\documentclass[runningheads]{llncs}
\usepackage{epsfig}
\usepackage{graphicx}
\usepackage{amsmath}
\usepackage{amssymb}
\usepackage{enumitem}
\usepackage[belowskip=-12pt,aboveskip=-6pt]{caption}
\usepackage[pagebackref=true,breaklinks=true,colorlinks,bookmarks=false]{hyperref}
\usepackage{makecell}
\usepackage{array}
\usepackage{cite}
\usepackage[toc, page]{appendix}
\usepackage{subcaption,siunitx,booktabs}
\usepackage[section]{placeins}
\begin{document}
\title{Deformable Cross-Attention Transformer for Medical Image Registration}
%
%
\author{Junyu Chen \inst{1} Yihao Liu \inst{2} \and Yufan He \inst{3} \and Yong Du \inst{1} \\\email{\{jchen245,yliu236,duyong\}@jhmi.edu;yufanh@nvidia.com}}
%
\authorrunning{J. Chen et al.}
%
\institute{Russell H. Morgan Department of Radiology and Radiological Science,\\ Johns Hopkins Medical Institutes, Baltimore, MD, USA \and Department of Electrical and Computer Engineering,\\ Johns Hopkins University, Baltimore, MD, USA \and NVIDIA Corporation, Bethesda, MD, USA}
\maketitle              
\begin{abstract}
Transformers have recently shown promise for medical image applications, leading to an increasing interest in developing such models for medical image registration. Recent advancements in designing registration Transformers have focused on using cross-attention (CA) to enable a more precise understanding of spatial correspondences between moving and fixed images. Here, we propose a novel CA mechanism that computes windowed attention using deformable windows. In contrast to existing CA mechanisms that require intensive computational complexity by either computing CA globally or locally with a fixed and expanded search window, the proposed deformable CA can selectively sample a diverse set of features over a large search window while maintaining low computational complexity. The proposed model was extensively evaluated on multi-modal, mono-modal, and atlas-to-patient registration tasks, demonstrating promising performance against state-of-the-art methods and indicating its effectiveness for medical image registration. The source code for this work will be available after publication.

\end{abstract}
\section{Introduction}
Deep learning-based registration methods have emerged as a faster alternative to optimization-based methods, with promising registration accuracy across a range of registration tasks~\cite{balakrishnan2019voxelmorph, kim2021cyclemorph}. 
These methods often adopt convolutional neural networks (ConvNets), particularly U-Net-like networks~\cite{ronneberger2015u}, as the backbone architecture \cite{balakrishnan2019voxelmorph, kim2021cyclemorph}.
Yet, due to the locality of the convolution operations, the effective receptive fields (ERFs) of ConvNets are only a fraction of their theoretical receptive fields~\cite{li2022transforming, luo2016understanding}.
This limits the performance of ConvNets in image registration, which often requires registration models to establish long-range spatial correspondences between images.

\begin{figure}[t]
\begin{center}
\includegraphics[width=0.6\textwidth]{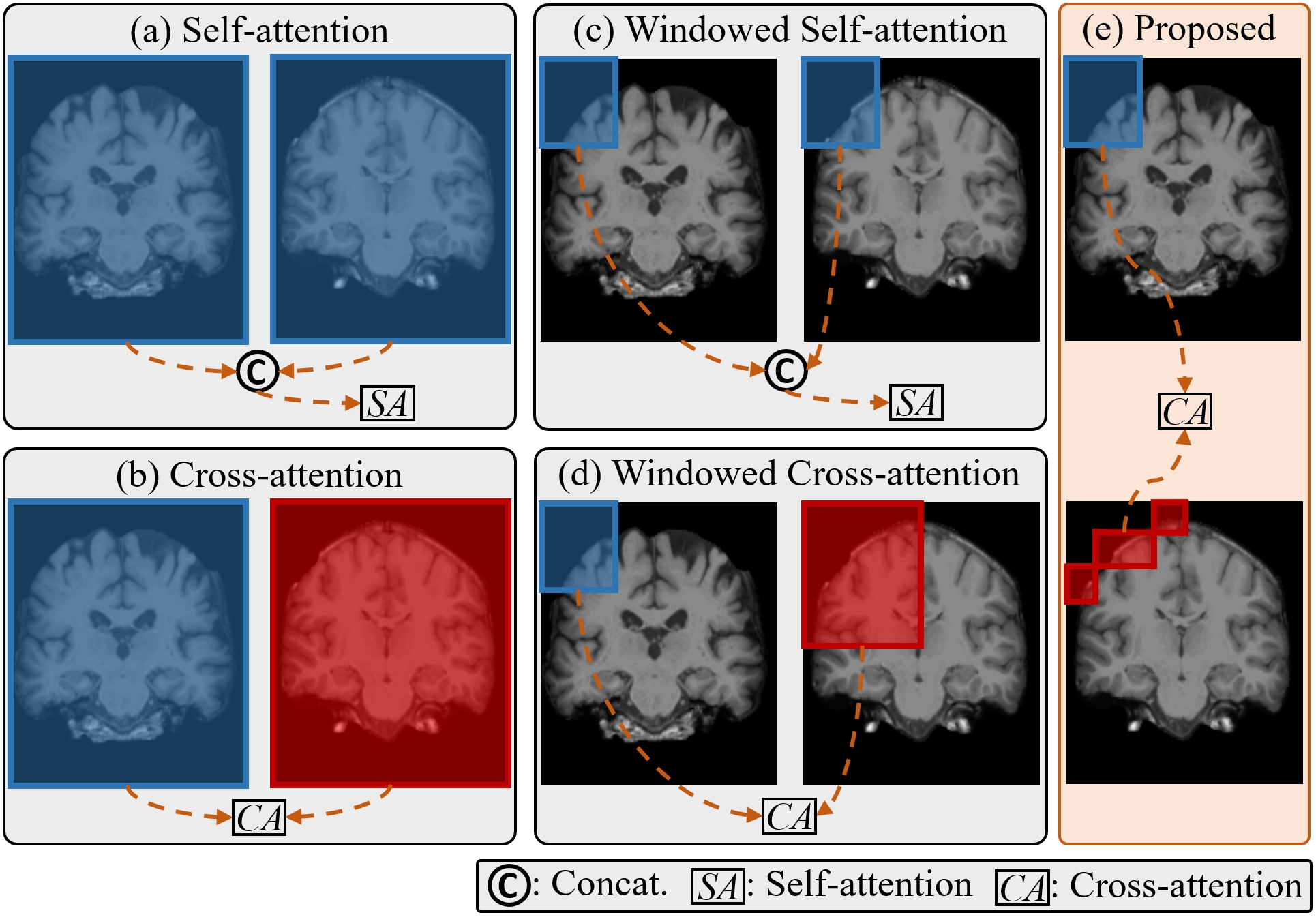}
\end{center}
   \caption{Graphical illustrations of different attention mechanisms. (a) The conventional self-attention~\cite{vaswani2017attention, dosovitskiy2020image} used in \texttt{ViT-V-Net}~\cite{chen2021vitvnet} and \texttt{DTN}~\cite{zhang2021learning}, which computes attention for the concatenated tokens of moving and fixed images. (b) Cross-attention used in \texttt{Attention-Reg}~\cite{song2022cross}, which computes attention between the tokens of moving and fixed images. (c) Windowed self-attention~\cite{liu2021swin} used in \texttt{TransMorph}~\cite{chen2022transmorph}, which computes attention for the concatenated tokens of moving and fixed images within a local window. (d) Windowed cross-attention proposed in \texttt{XMorpher}~\cite{shi2022xmorpher}, which computes attention between the tokens of fixed and moving images, specifically between two local windows of different sizes. (e) The proposed deformable cross-attention mechanism, which computes attention between tokens within a rectangular window and a deformed window with an arbitrary shape but the same size as the rectangular window.}
\label{fig:cross_attn}
\end{figure}

Transformers, which originated from natural language processing tasks~\cite{vaswani2017attention}, have shown promise in a variety of medical imaging applications \cite{li2022transforming}, including registration \cite{chen2021vitvnet, chen2022transmorph, zhang2021learning}. 
Transformers employ the self-attention (SA) mechanism, which can either be a global operation~\cite{dosovitskiy2020image} or computed locally within large windows~\cite{liu2021swin}.
Consequently, Transformers have been shown to capture long-range spatial correspondences for registration more effectively than ConvNets~\cite{chen2022transmorph}.

Several recent advancements in Transformer-based registration models have focused on developing cross-attention (CA) mechanisms, such as \texttt{XMorpher}~\cite{shi2022xmorpher} and \texttt{Attention-Reg}~\cite{song2022cross}.
CA improves upon SA by facilitating the efficient fusion of high-level features between images to improve the comprehension of spatial correspondences.
However, the existing CA mechanisms still have drawbacks; either they compute CA globally~\cite{song2022cross}, which prevents hierarchical feature extraction and applies only to low-resolution features, or they compute CA within a fixed but expanded window~\cite{shi2022xmorpher}, which significantly increases computational complexity.

In this paper, we present a hybrid Transformer-ConvNet model based on a novel deformable CA mechanism for image registration. 
As shown in Fig. \ref{fig:cross_attn}, the proposed deformable CA module differs from existing SA and CA modules in that it employs the windowed attention mechanism~\cite{liu2021swin} with a learnable offset. This allows the sampling windows of the reference image to take on any shapes based on the offsets, offering several advantages over existing methods:
\textbf{1)} In contrast to the CA proposed in \cite{shi2022xmorpher}, which calculates attention between windows of varying sizes, the proposed deformable CA module samples tokens from a larger search region, which can even encompass the entire image.
Meanwhile, the attention computation is confined within a uniform window size, thereby keeping the computational complexity low.
\textbf{2)} The deformable CA enables the proposed model to focus more on the regions where the disparity between the moving and fixed images is significant, in comparison to the baseline ConvNets and SA-based Transformers, leading to improved registration performance.
Comprehensive evaluations were conducted on mono- and multi-modal registration tasks using publicly available datasets. The proposed model competed favorably against existing state-of-the-art methods, showcasing its promising potential for a wide range of image registration applications.

\section{Background and Related Works}
\label{sec:existing_CA}
\noindent\textbf{Self-attention.} SA~\cite{vaswani2017attention, dosovitskiy2020image} is typically applied to a set of tokens (\emph{i.e.}, embeddings that represent patches of the input image). Let $\pmb{x}\in\mathbb{R}^{N\times D}$ be a set of $N$ tokens with $D$-dimensional embeddings. The tokens are first encoded by a fully connected layer $\pmb{U}_{q,k,v}\in\mathbb{R}^{D\times D_{q,k,v}}$ to obtain three matrix representations, Queries $\pmb{Q}$, Keys $\pmb{K}$, and Values $\pmb{V}$: $[\pmb{Q},\pmb{K},\pmb{V}] = \pmb{x}\pmb{U}_{q,k,v}$. Subsequently, the scaled dot-product attention is calculated using $SA(\pmb{x}) = \text{softmax}(\frac{\pmb{Q}\pmb{K}^\top}{\sqrt{D_k}})\pmb{V}.$
In general, SA computes a normalized score for each token based on the dot product of $\pmb{Q}$ and $\pmb{K}$. This score is then used to decide which Value token to attend to. 
\begin{figure}[t]
\begin{center}
\includegraphics[width=0.8\textwidth]{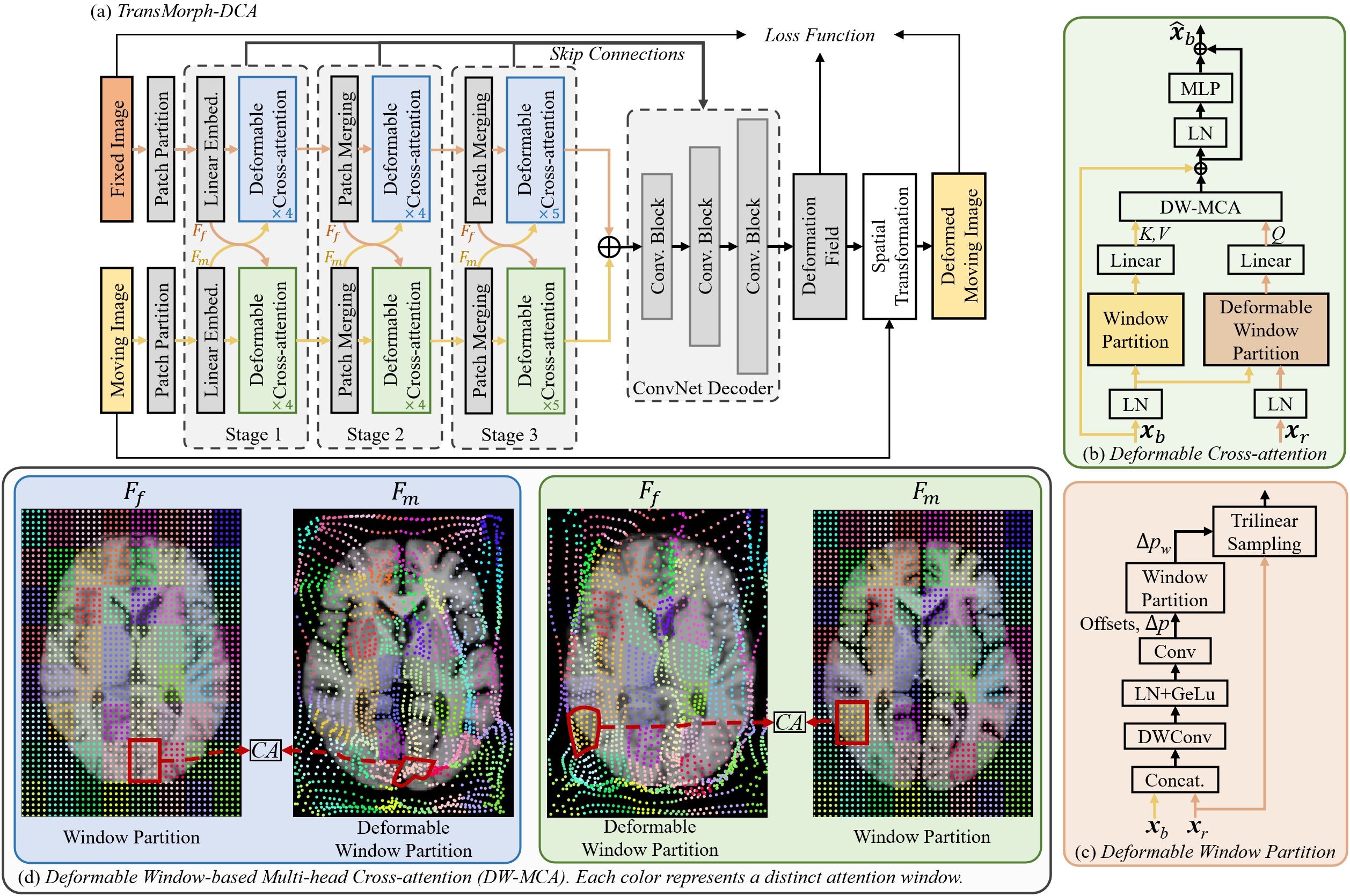}
\end{center}
   \caption{The overall framework of the proposed method. (a) The proposed network architecture, which is composed of parallel Transformer encoders and a ConvNet decoder to generate a deformation field. (b) The deformable CA, which fuses features between encoders. (c) The schematic of the deformable window partitioning strategy. (d) An example of deformable CA computation in the DW-MCA.}
\label{fig:overall}
\end{figure}

\noindent\textbf{Cross-attention.} CA is a frequently used variant of SA for inter- and intra-modal tasks in computer vision~\cite{xu2020cross, chen2021crossvit, kim2022cross} and has been investigated for its potential in image registration~\cite{shi2022xmorpher, song2022cross, liu2022coordinate}. CA differs from SA in terms of how the matrix representations are computed. As CA is typically used between two modalities or images (\emph{i.e.}, a base image and a reference image), the matrices $\pmb{Q}$, $\pmb{K}$, and $\pmb{V}$ are generated using different inputs:
\begin{equation}
\label{eqn:CA}
[\pmb{K}_b,\pmb{V}_b] = \pmb{x}_b\pmb{U}_{k,v},\ \ \ \pmb{Q}_r = \pmb{x}_r\pmb{U}_{q},\ \ \
CA(\pmb{x}) = \text{softmax}(\frac{\pmb{Q}_r\pmb{K}_b^\top}{\sqrt{D_k}})\pmb{V}_b,
\end{equation}
where $\pmb{x}_b$ and $\pmb{x}_r$ denote, respectively, the tokens of the base and the reference.
In \cite{song2022cross}, Song \emph{et al.} introduced \texttt{Attention-Reg}, which employs Eqn. \ref{eqn:CA} to compute CA between a moving and a fixed image.
To ensure low computational complexity, CA is computed globally between the downsampled features extracted by ConvNets. 
However, because CA is only applied to a single resolution, it does not provide hierarchical feature fusion across different resolutions, a factor that is deemed important for several successful registration models~\cite{mok2022affine, mok2020large}. 
More recently, Shi \emph{et al.} introduced \texttt{XMorpher}~\cite{shi2022xmorpher}, which is based on the Swin Transformer~\cite{liu2021swin}. In \texttt{XMorpher}, CA is computed between the local windows of the tokens of different resolutions, enabling hierarchical feature fusion. As shown in Fig. \ref{fig:cross_attn} (d), the local windows are of different sizes, with a base window of size $N_b=h\times w\times d$ and a larger search window of size $N_s=\alpha h\times \beta w\times \gamma d$, where $\alpha$, $\beta$, and $\gamma$ are set equally to 3. Using a larger search window facilitates the effective establishment of spatial correspondence, but it also increases the computational complexity of each CA module. Specifically, if the same window size of $N_b$ is used, the complexity of CA is approximately $O(N_b^2D_k)$. However, if windows of different sizes, $N_b$ and $N_s$, are used, the complexity becomes $O(N_bN_sD_k)=O(\alpha\beta\gamma N_b^2D_k)$, where $\alpha\beta\gamma=3\times3\times3=27$. This means that using a larger search window would increase the computational complexity dramatically (by 27 times) and quickly become computationally infeasible.

Enlarging the search space while keeping computational costs low is challenging for 3D medical image registration. In this paper, we try to solve it with a deformable CA module that operates on equal-sized windows. This module not only provides hierarchical feature fusion, but also allows more efficient token sampling over a larger region than previously mentioned CA modules. Additionally, the proposed CA maintains a low computational complexity.

\section{Proposed Method}
The proposed model is depicted in Fig. \ref{fig:overall} (a), which has dual Transformer encoders with deformable CA modules that enable effective communication between them. 
Each encoder is similar to the Swin~\cite{liu2021swin} used in \texttt{TransMorph}~\cite{chen2022transmorph}, but the SA modules are replaced with the deformable CA modules. 
To integrate the features between each stage of the two encoders, we followed \cite{song2022cross} by adding the features and passing them to the decoder via skip connections. 
In contrast to \texttt{XMorpher}~\cite{shi2022xmorpher}, which uses a Transformer for the decoder, we opted for the ConvNet decoder introduced in \cite{chen2022unsupervised, chen2022transmorph}. 
This choice was motivated by the inductive bias that ConvNets bring in, which Transformers typically lack~\cite{li2022transforming}. 
ConvNets are also better at refining features for subsequent deformation generation, owing to the locality of convolution operations. 
Moreover, ConvNets have fewer parameters, making them efficient and hence speeding up the training process.

\noindent{\textbf{Cross-attention Transformer.}} Our model employs parallel deformable CA encoders to extract hierarchical features from the moving and fixed images in the encoding stage. At each resolution of the encoder, $k$ successive deformable CA modules are applied to vertically fuse features between the two encoders. The deformable cross-attention module takes in a base (\emph{i.e.}, $\pmb{x}_b$) and a reference (\emph{i.e.}, $\pmb{x}_r$), and computes the attention between them, with the reference guiding the network on where to focus within the base. As shown in Fig. \ref{fig:overall} (a), one encoding path uses the moving and fixed images as the base and reference, respectively, whereas the other encoding path switches the roles of the base and reference, using the moving image as the reference and the fixed image as the base.

\noindent{\textbf{Deformable Cross-attention.}} Fig. \ref{fig:overall} (b) depicts the core element of the proposed model, the deformable CA. The module first applies \textit{LayerNorm} (LN) to $\pmb{x}_b$ and $\pmb{x}_r$, then  partitions $\pmb{x}_b$ into non-overlapping rectangular equal-sized windows, following \cite{liu2021swin}. Next, the $\pmb{x}_b$ is projected into $\pmb{K}_b$ and $\pmb{V}_b$ embeddings through a linear layer. This process is expressed as $[\pmb{K}_b,\pmb{V}_b] = \text{WP}(\text{LN}(\pmb{x}_b))\pmb{U}_{k,v}$, where $\text{WP}(\cdot)$ denotes the window partition operation. 
On the other hand, the window partitioning for $\pmb{x}_r$ is based on the offsets, $\Delta p$, learned by a lightweight offset network. As shown in Fig. \ref{fig:overall} (c), this network comprises two consecutive convolutional layers (depth-wise and regular convolutional layers) and takes the added $\pmb{x}_b$ and $\pmb{x}_r$ as input. The offsets, $\Delta p$, shift the sampling positions of the rectangular windows beyond their origins, allowing tokens to be sampled outside these windows. Specifically, $\Delta p$ are first divided into equal-sized windows, $\Delta p_w$, and tokens in $\pmb{x}_r$ are subsequently sampled based on $\Delta p_w$ using trilinear interpolation. Note that this sampling process is analogous to first resampling the tokens based on the offsets and then partitioning them into windows. We generated a different set of $\Delta p_w$ for each head in the multi-head attention, thereby enabling diverse sampling of the tokens across heads. The proposed deformable window-based multi-head CA (DW-MCA) is then expressed as:
\begin{equation}
\label{eqn:def_CA}
\begin{split}
    [\pmb{K}_b,\pmb{V}_b] &= \text{WP}(\text{LN}(\pmb{x}_b))\pmb{U}_{k,v},\\
   \Delta p = \theta_{\Delta p}(\pmb{x}_b, \pmb{x}_r),\ \ \Delta p_w &=\text{WP}(\Delta p), \ \ \pmb{Q}_r = \psi(\pmb{x}_r; p+\Delta p_w)\pmb{U}_{k},\\
    \text{DW-MCA}(\pmb{x}) &= \text{softmax}(\frac{\pmb{Q}_r\pmb{K}_b^\top}{\sqrt{D_k}})\pmb{V}_b,
\end{split}
\end{equation}
where $\theta_{\Delta p}$ denotes the offset network and $\psi(\cdot;\cdot)$ is the interpolation function. To introduce cross-window connections, the shifted window partitioning strategy~\cite{liu2021swin} was implemented in successive Transformer blocks.

The attention computation of the deformable CA is nearly identical to the conventional window-based SA employed in Swin~\cite{liu2021swin}, with the addition of a lightweight offset network whose complexity is approximately $O(m^3N_bD_k)$ ($m$ is the convolution kernel size and $m^3\approx N_b$).
As a result, the overall complexity of the proposed CA module is $O(2N_b^2D_k)$, which comprises the complexity of the offset network and the CA computation.
In comparison, the CA used in \texttt{XMorpher}~\cite{shi2022xmorpher} has a complexity of $O(27N_b^2D_k)$, as outlined in section \ref{sec:existing_CA}.
This highlights the three main advantages of the deformable CA module: \textbf{1)} it enables token sampling beyond a pre-defined window, theoretically encompassing the entire image size, \textbf{2)} it allows sampling windows to overlap, improving communication between windows, and \textbf{3)} it maintains fixed-size windows for the CA computation, thereby retaining low computational complexity.

The deformable CA, the deformable attention (DA)~\cite{xia2022vision}, and the Swin DA (SDA)~\cite{huang2022swin} share some similarities, but there are fundamental differences. Firstly, DA computes attention globally within a single modality or image, whereas the deformable CA utilizes windowed attention and a hierarchical architecture to fuse features of different resolutions across images or modalities. Secondly, the offset network in DA and SDA is applied solely to the Query embeddings of input tokens, and SDA generates offsets based on window-partitioned tokens, leading to square-shaped ``windowing" artifacts in the sampling grid, as observed in \cite{huang2022swin}. In contrast, in the deformable CA, the offset network is applied to all tokens of both the reference and the base to take advantage of their spatial correspondences, resulting in a smoother and more meaningful sampling grid, as demonstrated in Figure \ref{fig:overall} (d). Lastly, while DA and SDA use a limited number of reference points to interpolate tokens during sampling, deformable CA employs a dense set of reference points with the same resolution as the input tokens, allowing deformable CA to sample tokens more diversely.

\section{Experiments}
\noindent{\textbf{Dataset and Pre-processing.}} The proposed method was tested on three publicly available datasets to evaluate its performance on three registration tasks: \textbf{1)} inter-patient multi-modal registration, \textbf{2)} inter-patient mono-modal registration, and \textbf{3)} atlas-to-patient registration.
The dataset used for the first task is the ALBERTs dataset~\cite{gousias2012magnetic}, which consists of T1- and T2-weighted brain MRIs of 20 infants. Manual segmentation of the neonatal brain was provided, each consisting of 50 ROIs. The patients were randomly split into three sets with a ratio of 10:4:6. We performed inter-patient T1-to-T2 registration, which resulted in 90, 12, and 30 image pairs for training, validation, and testing, respectively. For the second and third registration tasks, we used the OASIS dataset~\cite{marcus2007open} from the Learn2Reg challenge~\cite{hering2022learn2reg} and the IXI dataset\footnote{https://brain-development.org/ixi-dataset/} from \cite{chen2022transmorph}, respectively. The former includes 413 T1 brain MRI images, of which 394 were assigned for training and 19 for testing. The latter consists of 576 T1 brain MRI images, which were distributed as 403 for training, 58 for validation, and 115 for testing. For the third task, we used a moving image, which was a brain atlas image obtained from \cite{kim2021cyclemorph}. All images from the three datasets were cropped to the dimensions of $160 \times 192 \times 224$.


\noindent{\textbf{Evaluation Metrics.}} To assess the registration performance, the Dice coefficient was used to measure the overlap of the anatomical label maps. Moreover, for the OASIS dataset, we additionally used Hausdorff distance (HdD95) to evaluate performance and the standard deviation of the Jacobian determinant (SDlogJ) to assess deformation invertibility, in accordance with Learn2Reg. For the ALBERTs and IXI datasets, we used two metrics, the percentage of all non-positive Jacobian determinant (\%$\vert J\vert\leq0$) and the non-diffeomorphic volume (\%NDV), both proposed in \cite{liu2022finite}, to evaluate deformation invertibility since they are more accurate measures under the finite-difference approximation.

\begin{table}[t]
\centering
\begin{subtable}{1\textwidth}
\fontsize{5.5}{7}\selectfont
    \begin{tabular}{ c | c c c m{0.0001\textwidth} c | c c c}
 \Xhline{1pt}
 \multicolumn{4}{c}{\textbf{OASIS} (Mono-modality)}& &\multicolumn{4}{c}{\textbf{IXI} (Atlas-to-patient)}\\
 \cline{1-4}\cline{6-9}
 Method & Dice$\uparrow$ & HdD95$\downarrow$ & SDlogJ$\downarrow$ && Method & Dice$\uparrow$ & \%$\vert J\vert\leq0\downarrow$ & \%NDV$\downarrow$ \\
 \cline{1-4}\cline{6-9}
 ConvexAdam~\cite{siebert2021fast} & 0.846$\pm$0.016 & 1.500$\pm$0.304 & 0.067$\pm$0.005 && VoxelMorph~\cite{balakrishnan2019voxelmorph} & 0.732$\pm$0.123&6.26\%&1.04\%\\
\cline{1-4}\cline{6-9}
 LapIRN~\cite{mok2021conditional} & 0.861$\pm$0.015 & 1.514$\pm$0.337 & 0.072$\pm$0.007 && CycleMorph~\cite{kim2021cyclemorph} & 0.737$\pm$0.123&6.38\%&1.15\%\\
 \cline{1-4}\cline{6-9}
 TransMorph~\cite{chen2022transmorph} & 0.862$\pm$0.014& 1.431$\pm$0.282 & 0.128$\pm$0.021  && TM-bspl~\cite{chen2022transmorph} & 0.761$\pm$0.128 & 0\% &0\%\\
 \cline{1-4}\cline{6-9}
 TM-TVF~\cite{chen2022unsupervised} & \textit{0.869$\pm$0.014}& \textbf{1.396$\pm$0.295} & 0.094$\pm$0.018 && TM-TVF~\cite{chen2022unsupervised} & 0.756$\pm$0.122 & 2.05\% & 0.36\% \\
 \cline{1-4}\cline{6-9}
 XMorpher*~\cite{shi2022xmorpher} & 0.854$\pm$0.012 & 1.647$\pm$0.346 & 0.100$\pm$0.016 && XMorpher*~\cite{shi2022xmorpher} & 0.751$\pm$0.123 & 0\% & 0\% \\
 \Xhline{1pt}
 TM-DCA & \textbf{0.873$\pm$0.015}& \textit{1.400$\pm$0.368} & 0.105$\pm$0.028 && TM-DCA & \textbf{0.763$\pm$0.128} & 0\% & 0\%\\
 \cline{1-4}\cline{6-9}
 \Xhline{1pt}
\end{tabular}
\end{subtable}

\begin{subtable}{1\textwidth}
\centering
\fontsize{5.5}{7}\selectfont
    \begin{tabular}{ c | c c c }
 \Xhline{1pt}
 \multicolumn{4}{c}{\textbf{ALBERTs} (Multi-modality)}\\
 \cline{1-4}
 Method & Dice$\uparrow$ & \%$\vert J\vert\leq0\downarrow$ & \%NDV$\downarrow$\\
 \cline{1-4}
 VoxelMorph~\cite{siebert2021fast} & 0.651$\pm$0.159 & 0.04\% & 0.02\% \\
\cline{1-4}
 TransMorph~\cite{chen2022transmorph} & 0.672$\pm$0.159 & 0.15\% & 0.04\%\\ 
 \cline{1-4}
 TM-TVF~\cite{chen2022unsupervised} & \textit{0.722$\pm$0.132}  & 0.13\% & 0.03\% \\
 \cline{1-4}
 XMorpher*~\cite{shi2022xmorpher} & 0.710$\pm$0.135 & 0.11\% & 0.03\%\\
 \Xhline{1pt}
 TM-DCA & \textbf{0.724$\pm$0.131} & 0.24\% & 0.07\%\\
 \cline{1-4}
 \Xhline{1pt}
\end{tabular}
\end{subtable}
\caption{Quantitative results for mono-modal (OASIS) and multi-modal inter-patient (ALBERTs) registration tasks, as well as atlas-to-patient (IXI) registration tasks. Note that part of the OASIS results was obtained from Learn2Reg leaderboard~\cite{hering2022learn2reg}.}\label{tab:oasis_ixi_alberts}
\end{table}

\begin{figure}[t]
\begin{center}
\includegraphics[width=1.\textwidth]{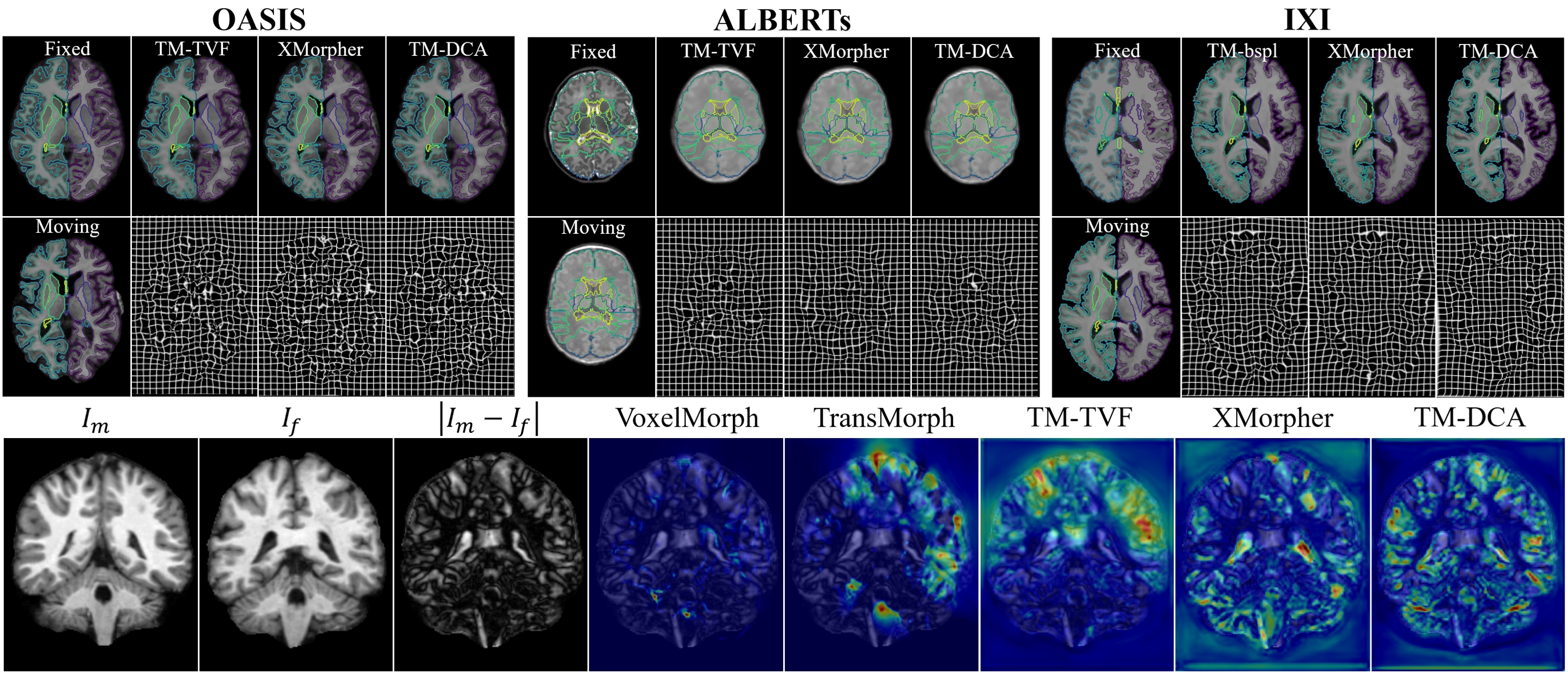}
\end{center}
   \caption{Qualitative results and a visualization of Grad-CAM~\cite{selvaraju2017grad} heat maps of the comparative registration models generated for a pair of images.}
\label{fig:gradcam}
\end{figure}

\noindent{\textbf{Results and Discussion.}}
The proposed model, \texttt{TM-DCA}, was evaluated against several state-of-the-art models on the three registration tasks, and the corresponding quantitative outcomes are presented in Table \ref{tab:oasis_ixi_alberts}. Note that our GPU was unable to accommodate the original \texttt{XMorpher}~\cite{shi2022xmorpher} ($>$48 GB) for the image size used in this study.
This is likely due to the large window CA computation and the full Transformer architecture used by the model. To address this, we used the encoder of \texttt{XMorpher} in combination with the decoder of \texttt{TM-DCA} (denoted as \texttt{XMorpher}*) to reduce GPU burden and facilitate a more precise comparison between the deformable CA in \texttt{TM-DCA} and the CA used in \texttt{XMorpher}.
On the OASIS dataset, \texttt{TM-DCA} achieved the highest mean Dice score of 0.873, which was significantly better than the second-best performing method, \texttt{TM-TVF}~\cite{chen2022unsupervised}, as confirmed by a paired t-test with $p<0.01$. On the IXI dataset, \texttt{TM-DCA} achieved the highest mean Dice score of 0.763, which was significantly better than \texttt{TM-bspl} with $p<0.01$.
Remarkably, \texttt{TM-DCA} produced diffeomorphic registration with almost no folded voxels, using the same decoder as \texttt{TM-bspl}.
Finally, on the ALBERTs dataset, \texttt{TM-DCA} again achieved the highest mean Dice score of 0.713, which was significantly better than TM-TVF with $p=0.04<0.05$, thus demonstrating its superior performance in multi-modal registration.
It is important to note that the proposed \texttt{TM-DCA} model and its CA (\emph{i.e.}, \texttt{XMorpher}) and SA (\emph{i.e.}, \texttt{TM-TVF} and \texttt{TM-bspl}) counterparts differed only in their encoders, while the decoder used was identical for all models. \texttt{TM-DCA} consistently outperformed the baselines across the three applications, supporting the effectiveness of the proposed CA module.

Qualitative comparison results between the registration models are presented in Fig.~\ref{fig:gradcam}. In addition, we conducted a comparison of the \textit{Grad-CAM}~\cite{selvaraju2017grad} heat map for various learning-based registration models. The heat maps were generated by computing the NCC between the deformed moving image and the fixed image, and were then averaged across the convolutional layers at the end of the decoder, just prior to the final layer that predicts the deformation field. Notably, \texttt{VoxelMorph} exhibited inadequate focus on the differences between the image pair, which may be attributed to ConvNets' limited ability to explicitly comprehend contextual information in the image. The SA-based models (\texttt{TransMorph} and \texttt{TM-TVF}) showed similar trends, wherein they focused reasonably well on regions with significant differences but relatively less attention was given to areas with minor differences. In contrast, the attention of CA-based models was more uniformly distributed, with the proposed \texttt{TM-DCA} method more effectively capturing differences than \texttt{XMorpher}. The presented heat maps highlight the superior performance of the proposed CA mechanism in effectively interpreting contextual information and accurately capturing spatial correspondences between images. In combination with the observed improvements in performance across various registration tasks, these results suggest that \texttt{TM-DCA} has significant potential as the preferred attention mechanism for image registration applications.

\section{Conclusion}
In this study, we introduced a Transformer-based network for unsupervised image registration. The proposed model incorporates a novel CA module that computes attention between the features of the moving and fixed images. Unlike the SA and CA mechanisms used in existing methods, the proposed CA module computes attention between tokens sampled from a square window and a learned window of arbitrary shape. This enables the efficient computation of attention while allowing the extraction of useful features from a large window to accurately capture spatial correspondences between images. The proposed method was evaluated against several state-of-the-art methods on multiple registration tasks and demonstrated significant performance improvements compared to the baselines, highlighting the effectiveness of the proposed CA module.

%
%
%
\bibliographystyle{splncs04}
\bibliography{mybibliography}

\section*{Appendix}
\appendix
\section{Data Preprocessing}
\noindent\textbf{ALBERTs.} We used a kernel density estimate-based method to normalize the images. Additionally, since patients exhibit considerable anatomical differences due to rapid growth, we performed affine registration to align all patients affinely with the first patient. This approach ensures that the only cause of misalignment among the volumes is nonlinear.

\noindent\textbf{IXI \& OASIS.} We used FreeSurfer to perform standard procedures for brain MRI. Anatomical label maps, including over 30 ROIs, were generated for both datasets. 

\section{Hyperparameters Settings}
\label{hyper}
\begin{table}[!hbp]
\centering
\fontsize{6.5}{8}\selectfont
    \begin{tabular}{ c | c c c c c c c c c c c c}
 \hline
     & Loss & Loss Wt. & Decoder & Data Aug. & Patch Sz.& Embd. Sz. & Layer Num. ($k$)\\
 \hline
 OASIS  & NCC+Dice+Diff. & $[1, 1, 1]$ & TM-TVF~\cite{chen2022unsupervised} & - & 4 & 96 & $\{4, 4, 5\}$ \\
 \hline
 IXI  & NCC+Diff. & $[1, 1]$ & TM-bspl~\cite{chen2022transmorph} & Rand. Flip & 4 & 96 & $\{4, 4, 5\}$ \\
 \hline
 ALBERTs  & MIND+Dice+Diff.  & $[1, 1, 1]$ & TM-TVF~\cite{chen2022unsupervised} & Rand. Affine & 4 & 96 & $\{4, 4, 5\}$\\
 \hline
\end{tabular}
\caption{Training setups for the proposed registration model. The proposed registration model was trained using the decoder from \texttt{TM-TVF} and \texttt{TM-bspl} for different datasets. The window size was set to $\{5, 6, 7\}$, which is consistent with the window sizes used in \texttt{TransMorph}. The time step used in \texttt{TM-TVF} was set to 7. The models were trained for 500 epochs with the Adam optimizer and a learning rate of 1$e$-4. The PyTorch framework was used for model implementation, and training was performed on an NVIDIA A6000 GPU.}\label{table_hyper}
\end{table}
\clearpage
\section{Deformable Window Partition \& Window Partition}
\begin{figure}[!htb]
\begin{center}
\includegraphics[width=1\textwidth]{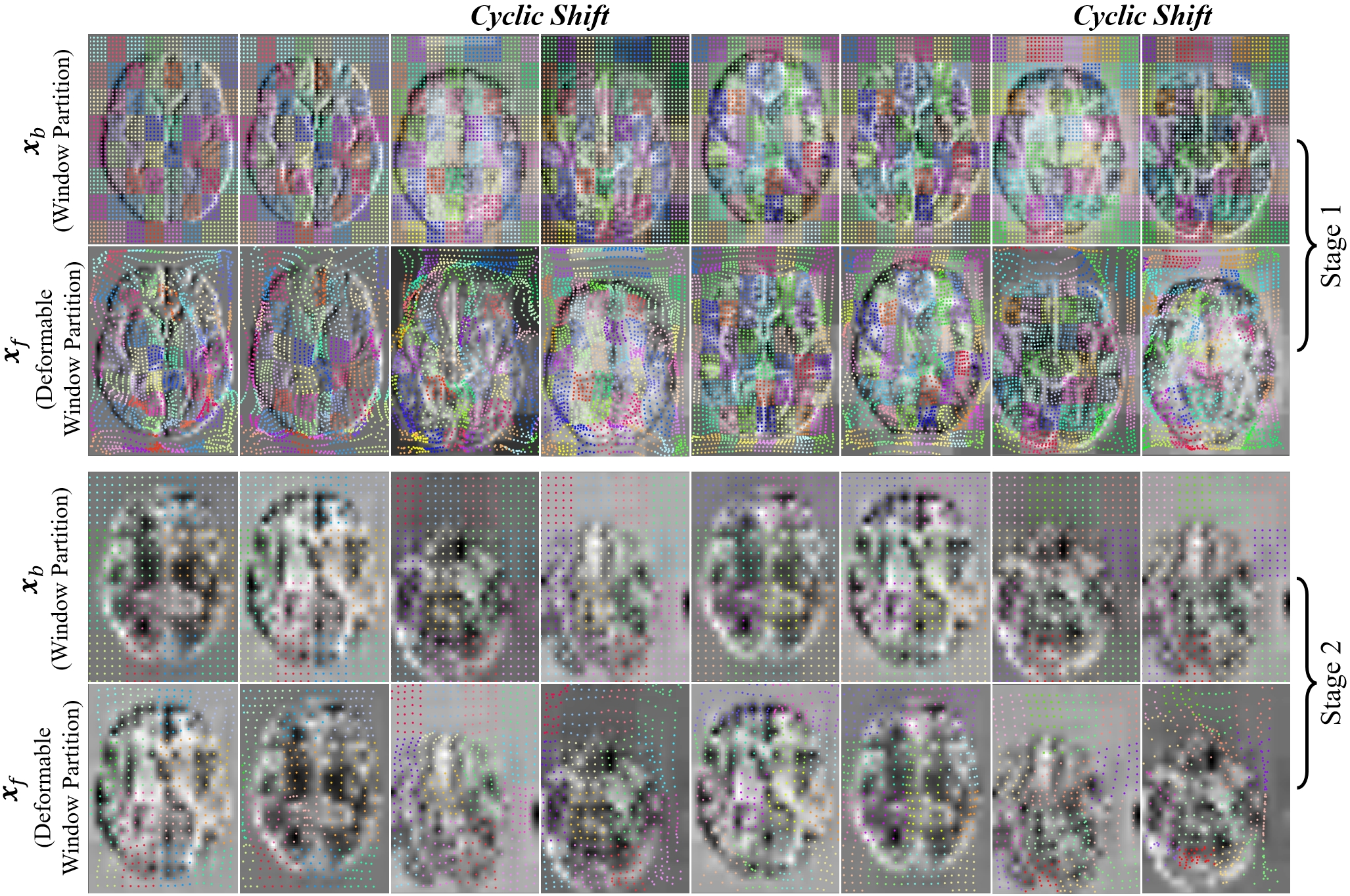}
\end{center}
   \caption{The proposed window partitioning strategy in the first and second stages of the network is visualized in the figure, where each color represents a unique sampling window. Recall that the deformable window partitioning is applied to the reference image $\pmb{x}_r$, while the rectangular window partitioning is applied to the base image $\pmb{x}_b$. Subsequently, the cross-attention is computed between the windows in $\pmb{x}_b$ and $\pmb{x}_r$. It is evident that the deformable window partitioning concentrates the sampling locations in information-rich regions.}
\label{fig:def_win_par}
\end{figure}
\clearpage
\section{Additional Quantitative Results}
\begin{figure}[!htb]
\begin{center}
\includegraphics[width=1\textwidth]{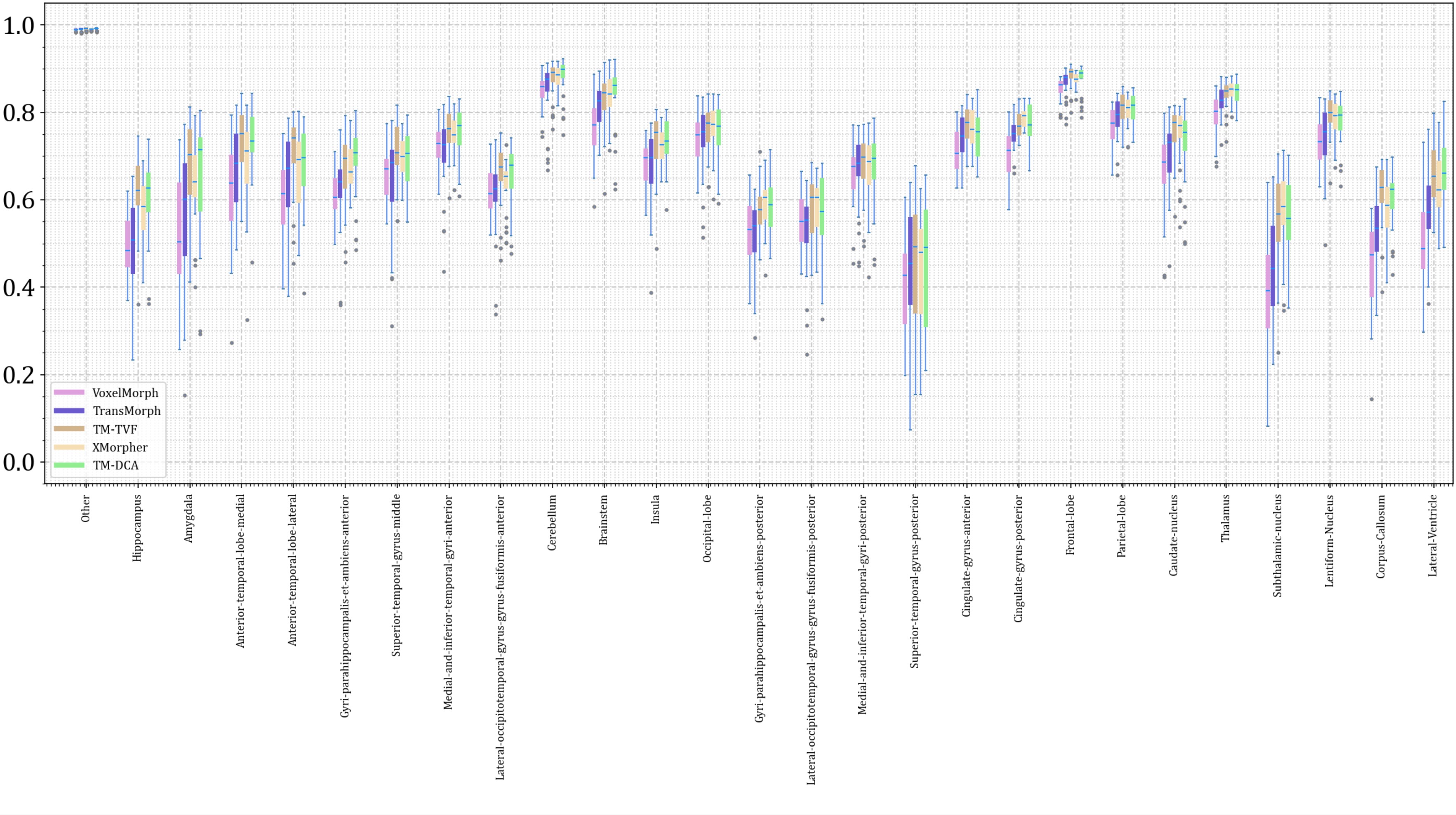}
\end{center}
   \caption{Quantitative results of multi-modal registration on the ALBERTs dataset.}
\label{fig:Albers_quant}
\end{figure}

\begin{figure}[!htb]
\begin{center}
\includegraphics[width=1\textwidth]{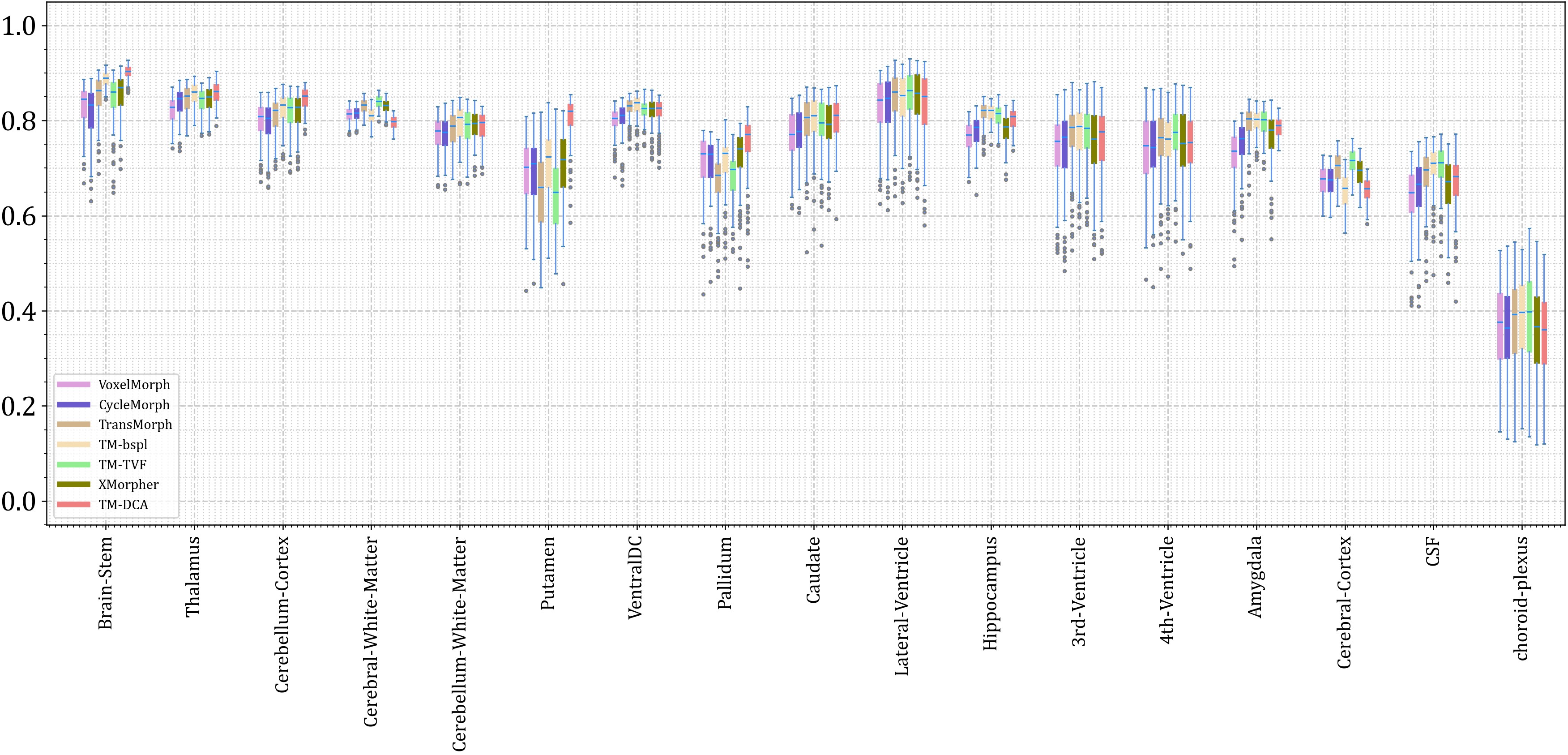}
\end{center}
   \caption{Quantitative results of atlas-to-patient registration on the IXI dataset.}
\label{fig:IXI_quant}
\end{figure}
\end{document}